# Parameter Estimation of Ground Moving Targets Based on SKT-DLVT Processing

Jing Tian, Wei Cui, and Si-liang Wu

*Abstract*—It is well known that the motion of a ground moving target may induce the range cell migration, spectrum spread and velocity ambiguity during the imaging time, which makes the image smeared. To eliminate the influence of these factors on image focusing, a novel method for parameter estimation of ground moving targets, known as SKT-DLVT, is proposed in this paper. In this method, the segmental keystone transform (SKT) is used to correct the range walk of targets simultaneously, and a new transform, namely, Doppler Lv's transform (LVT) is applied on the azimuth signal to estimate the parameters. Theoretical analysis confirms that no interpolation is needed for the proposed method and the targets can be well focused within limited searching range of the ambiguity number. The proposed method is capable of obtaining the accurate parameter estimates efficiently in the low signal-to-noise ratio (SNR) scenario with low computational burden and memory cost, making it suitable to be applied in memory-limited and real-time processing systems. The effectiveness of the proposed method is demonstrated by both simulated and real data.

*Index Terms*—ground moving target, segmental keystone transform (SKT), Doppler Lv's transform (LVT), parameter estimation.

I INTRODUCTION

Synthetic aperture radar (SAR) has been widely used in many civilian and military applications including moving target imaging and identification for its ability of information acquisition. The position shift of a moving target in a conventional SAR image cause the image defocusing; therefore, detection, parameter estimation, imaging and relocation of moving targets have received considerable attention in the radar imaging community [1-3].

For a SAR system with ground moving-target indication (GMTI) [4, 5], radar motion and long dwell time produce the large azimuth bandwidth which can be used to improve the signal-to-noise ratio (SNR). However, large range cell migration (RCM), spectrum spread and velocity ambiguity may occur [6-10], resulting in image defocusing when the conventional SAR imaging algorithms are applied to the observed scene with moving targets. Several methods have been proposed to refocus the moving target. These methods can be classified into two types. In the first type, targets should be detected before parameter estimation [11-14]. The image can be well focused with the estimates of motion parameters, which are achieved first by exploiting the range migration induced by the motion. However, these methods perform poorly in the case of the large RCM, spectrum spread or velocity ambiguity. In the second type, the moving targets can be well imaged without a priori knowledge of motion parameters. The motion parameters of targets can be estimated by optimizing the quality of the target image signature in [15]. During the focusing process, the RCM is completely corrected. However, it has a heavy computational burden when the parameter searching range is large. Keystone transform (KT) based methods have been proposed in [6][16][17]. However, these methods cannot correct the RCM completely in the case of velocity ambiguity, thereby impacting the energy integration and parameter estimation. A 2-D matched filtering method has been proposed in [4], which can correct the RCM without a priori knowledge of the accurate motion parameters. However, the azimuth defocusing may occur without knowing the along-track velocity information. An instantaneous-range-Doppler method based on deramp-keystone processing has been proposed in [9], which can focus a moving target at an arbitrarily chosen azimuth time without specific knowledge of its accurate motion parameters. This method can eliminate the RCM of multiple targets simultaneously and solve the problems of Doppler spectrum spread and velocity ambiguity. However, the position of the targets cannot be obtained directly and the Doppler spectrum spread cannot be compensated completely due to the mismatch of deramp function for the target with large azimuth velocity, which would further affect the precision of parameter estimates and azimuth focusing. The scaling processing and fractional Fourier transform (SPFRFT) method proposed in [18] can be applied for the compensation of RCM and Doppler spectrum spread, however, it has a heavy computational burden since it needs 3-dimensional searching. A new transform, i.e., stretch keystone-wigner transform (SKWT) has been proposed to estimate the motion parameters, which can resolve the RCM, velocity ambiguity and spectrum spread [19]. However, it suffers from heavy computational burden, bilinear transform and non-coherent integration.

In [20], a new parameter estimation method based on KT and Lv's transform (LVT) [21] has been proposed to reduce the computational burden. It has similar estimation precision to the fractional Fourier transform (FrFT), yet can be implemented without using any searching operation. For the linear frequency modulated (LFM) signals over long-time duration, however, the computational complexity and memory cost of LVT are huge, resulting in inaccessible requirement for DSP chips and unsuitable for real-time processing.

To deal with the problems of large RCM, Doppler spectrum spread and velocity ambiguity for the moving target, take the ability of processor, the data rate and the improvement of SNR into account, with as little priori knowledge as possible, this paper proposes a segmental keystone transform (SKT) and Doppler LVT based method (SKT-DLVT), which borrows the idea of segment to reduce the computational burden and storage memory cost. In this method, the SKT is used to correct the range walk and the Doppler LVT is applied on the azimuth signal to estimate the parameters. The major steps of the Doppler LVT include: 1) the fast Fourier transform (FFT) is applied on the azimuth signal within each segment; 2) the same frequency resolution bins of each segment are selected to construct new series; 3) Doppler KT is employed to correct the frequency walk across the segments; 4) inter-segment LVT is implemented to obtain the parameter estimates.

Unlike the conventional methods in [15-17, 22], the RCM correction of multiple moving targets in this paper is carried out simultaneously. The proposed estimator is accurate for the targets with Doppler spectrum spread and velocity ambiguity, which does not suffer from heavy computational burden by applying the parallel processing on the SKT and the Doppler LVT. In addition, the searching of the ambiguity number is only within a limited searching range. It is feasible, simple and suitable to be applied in memory-limited and real-time processing systems.

The remainder of this paper is organized as follows. Section II establishes the mathematical model of echo signal. Section III describes the proposed parameter estimation method for both slow and fast moving targets. In Section IV, some application considerations, such as the implementation of SKT, the criterion to choose the number of segments, the marginal velocity, the parameter estimation strategy for multiple moving targets, the output SNR, the computational complexity and memory cost, are analyzed in detail. Section V processes the simulated and real data to validate the proposed method. Section VI concludes the paper.

II SIGNAL MODELING

This section derives the signal model for a target moving with uniform rectilinear motion while ignores the higher order motion. The geometry relationship between the flying platform and the moving target is shown in Fig. 1, in which $V$, $v_a$ and $v_c$ denote the velocity of the platform, the along- and cross-track velocities of target, respectively. $R_B$ is the nearest range between the platform and the target, $t$ is the slow time. According to the geometry, the instantaneous slant range $R(t)$ between the platform and the target can be expressed as [4, 6]

$$R(t) = \sqrt{(Vt - v_a t)^2 + (R_B - v_c t)^2} \approx R_B - v_c t + (V - v_a)^2 t^2 / (2R_B) \qquad (1)$$

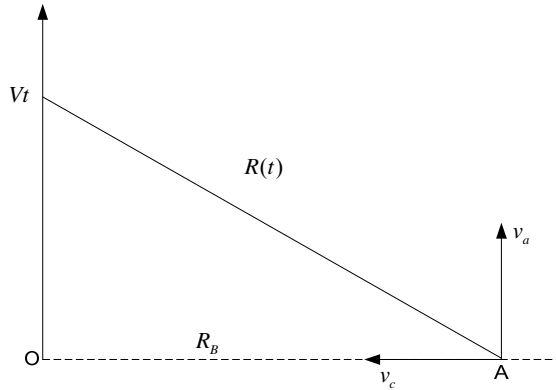

Fig.1. Geometry of moving target.

Assume the radar adopts LFM waveforms, i.e.,

$$s_T(t,\tau) = \text{rect}(\tau/T_p)\exp(j\pi\gamma\tau^2)\exp[j2\pi f_c(t+\tau)] \qquad (2)$$

where $\tau$ is the fast time, i.e., the range time; $t = nT(n = 0,1,\cdots N-1)$ is the slow time; $T$ is the pulse repetition time; $N$ is the number of coherent integrated pulses; rect(x) is the window function and equal to one for $|x| \leq 1/2$ or zero if otherwise; $T_p$ is the pulse width; $f_c$ is the carrier frequency; and $\gamma$ is the modulation rate. The received baseband signal after range compression can be expressed as [4]

$$s(t,\tau) = \sigma G w(t)\text{sinc}\left[\pi B\left(\tau - 2\frac{R(t)}{c}\right)\right]\exp\left[-j4\pi\frac{R(t)}{\lambda}\right] \qquad (3)$$

where $\sigma$ is the backscattering coefficient of the target, $G$ is the range compression gain, $w(t)$ is the azimuth window function [4], B is the bandwidth of the signal, c is the light speed, and $\lambda = c/f_c$ is the wavelength.

Substituting (1) into (3) yields

$$\begin{aligned}s(t,\tau) = \sigma G w(t)\text{sinc}&\left[\pi B\left(\tau - 2\frac{R_B - v_c t + (V-v_a)^2 t^2/(2R_B)}{c}\right)\right] \\ &\times \exp\left[-j4\pi\frac{R_B - v_c t + (V-v_a)^2 t^2/(2R_B)}{\lambda}\right]\end{aligned} \qquad (4)$$

Transforming $s(t,\tau)$ into the range-frequency and azimuth-time domain yields

$$S(t,f) = \frac{\sigma G w(t)}{B}\text{rect}\left(\frac{f}{B}\right)\exp\left\{-j4\pi\frac{f+f_c}{c}\left[R_B - v_c t + (V-v_a)^2 t^2/(2R_B)\right]\right\} \qquad (5)$$

It can be seen from (4) that the problems of the large RCM, Doppler spectrum spread and velocity ambiguity cannot concentrate the energy of the target completely, thereby making the image smeared. In the next section, we describe a new parameter estimation method. For the slow moving target without velocity ambiguity, this approach can obtain the estimates of targets without a priori knowledge of the motion information. For the fast moving target, i.e., in the presence of velocity ambiguity, this approach can estimate the parameters of targets with one-dimensional searching of the ambiguity number.

III METHOD FOR PARAMETER ESTIMATION

*A. Parameter Estimation Method for the Slow Moving Target*

We first present the algorithm for the target without velocity ambiguity [9], which satisfies $v \in [-\text{PRF}\lambda/4, \text{PRF}\lambda/4]$ with $\text{PRF}=1/T$. Since the large RCM of the target would affect the precision of parameter estimation, the correction of the large RCM should be implemented first using the proposed SKT.

The azimuth signal in (5) is firstly divided into segments with equal length, i.e., the azimuth time of $NT$ is divided into $P$ segments (the criterion to choose the number of segments is discussed in the Section IV-B later). Then substituting the scaling formula of the KT, i.e., $t_{seg} = \frac{f_c}{f+f_c} t_{a_{seg}}$ [23, 24], into the $seg$-th ($seg=1,2,...,P$) segment yields

$$
\begin{aligned}
S(t_{a_{seg}}, f) &= \frac{\sigma G w(t_{a_{seg}})}{B} \text{rect}\left(\frac{f}{B}\right) \\
&\times \exp\left\{-j4\pi \frac{f+f_c}{c}\left[R_B - v_c \frac{f_c}{f+f_c} t_{a_{seg}} + \frac{(V-v_a)^2}{2R_B}\left(\frac{f_c}{f+f_c} t_{a_{seg}}\right)^2\right]\right\} \\
&\approx \frac{\sigma G w(t_{a_{seg}})}{B} \text{rect}\left(\frac{f}{B}\right) \\
&\times \exp\left\{-j4\pi \frac{f+f_c}{c} R_B + j4\pi \frac{f_c}{c} v_c t_{a_{seg}} - j2\pi \frac{f_c}{c} \frac{(V-v_a)^2}{R_B}\left(1-\frac{f}{f_c}\right) t_{a_{seg}}^2\right\}
\end{aligned}
\quad (6)
$$

where $t_{a_{seg}} \in [(seg-1)NT/P : (seg-1)NT/P + (N/P-1)T]$.

Performing the inverse Fourier transform on $S(t_{a_{seg}}, f)$ with respect to $f$ yields

$$s(t_{a_{seg}},\tau) = \sigma G w(t_{a_{seg}})\mathrm{sinc}\left[\pi B\left(\tau - 2\frac{R_B - (V-v_a)^2 t_{a_{seg}}^2/(2R_B)}{c}\right)\right]$$
$$\times \exp\left[-j4\pi \frac{R_B - v_c t_{a_{seg}} + (V-v_a)^2 t_{a_{seg}}^2/(2R_B)}{\lambda}\right] \quad (7)$$

From (7), we find that the linear RCM has been removed completely. However, the quadratic RCM remains, which is related with $(V-v_a)^2 t_{a_{seg}}^2/(cR_B)$. This term has a minor influence on the RCM for C-band satellite SAR systems. However, for L-band satellites, the quadratic part is relatively large. In this situation, the quadratic RCM can be removed efficiently in the azimuth frequency domain [25]. After the quadratic RCM correction, the resulting signal is written as

$$s(t_{a_{seg}},\tau) = \sigma G w(t_{a_{seg}})\mathrm{sinc}\left[\pi B\left(\tau - 2\frac{R_B}{c}\right)\right]\exp\left[-j4\pi \frac{R_B - v_c t_{a_{seg}} + (V-v_a)^2 t_{a_{seg}}^2/(2R_B)}{\lambda}\right]$$
$$(8)$$

It can be seen from (8) that all the targets stay in the right range cells after range migration correction, which can improve the precision of estimation and further obtain the well-focused image. And it is obvious that the received signal from all scatters in one range cell can be modeled as a multi-component LFM signal after range compression and motion compensation. To obtain the accurate estimates of the velocities and accelerations of targets, we need to estimate the parameters of the LFM signal precisely. For simplicity, (8) can be further expressed as

$$x(t_{a_{seg}}) = \sigma_2 \exp\left[j2\pi\left(a_0 + a_1 t_{a_{seg}} + a_2 t_{a_{seg}}^2/2\right)\right] \quad (9)$$

where $\sigma_2 = \sigma G w(t_{a_{seg}})\mathrm{sinc}\left[\pi B\left(\tau - 2\frac{R_B}{c}\right)\right]$, $a_0 = -2R_B/\lambda$, $a_1 = 2v_c/\lambda$ and $a_2 = -2(V-v_a)^2/(\lambda R_B)$.

For multi-component LFM signals, the conventional time-frequency transform [26-30] suffers from performance degradation (even ineffective) because of the cross terms and the low-resolution problems in the low SNR scenario. These problems can be solved by applying the LVT for parameter estimation over the range cells. The LVT is able to obtain accurate parameter estimates without using any searching operation. This method breaks through the tradeoff between resolution and cross terms. For the LFM signals over long-time duration, however, the computational complexity and memory cost of LVT are huge, which would restrict its applications. Therefore, a new Doppler LVT method is proposed for parameter estimation. The core steps contain the segmental FFT processing of LFM signals and the inter-segment LVT applied on the new series constructed by the same frequency resolution bins of each segment.

Define $t_q = qT$ as the intra-segment time where $q = 0,1,...,N/P-1$ and $N/P$ is the number of samples within each segment, and define $t_p = (p-1)NT/P$ as the inter-segment time with $p = 1,2,...,P$. Then the azimuth time $t_{a_{seg}}$ is rewritten as $t_{a_{seg}} = t_q + t_p$. Ignoring the change of frequency within each segment interval, $x(t_{a_{seg}})$ can be approximated as

$$\begin{aligned}x(t_q,t_p) &= \sigma_2 \exp\left\{j2\pi\left[a_0 + a_1(t_q+t_p) + a_2(t_q+t_p)^2/2\right]\right\} \\ &\approx \sigma_2 \exp\left\{j2\pi\left[a_0 + a_1(t_q+t_p) + a_2(2t_q t_p + t_p^2)/2\right]\right\} \\ &= \sigma_2 \exp\left\{j2\pi\left[a_1 + a_2 t_p\right]t_q + j\varphi_p\right\}\end{aligned} \quad (10)$$

with $\varphi_p = 2\pi a_0 + 2\pi a_1 t_p + \pi a_2 t_p^2$.

The FFT of $x(t_q,t_p)$ with $t_q$ is computed to be

$$x(f_q, t_p) = \sigma_2 \frac{\sin\left\{\frac{N}{P}\pi\left[a_1 + a_2 t_p - f_q\right]\right\}}{\sin\left\{\pi\left[a_1 + a_2 t_p - f_q\right]\right\}} \exp(j\varphi_p)\exp\left[j\pi\left(\frac{N}{P}-1\right)(a_1 + a_2 t_p - f_q)\right] \quad (11)$$

It can be seen from (11) that the peak position of spectrum envelope varies with $t_p$ of each segment. And the frequency walk, which is larger than one frequency resolution bin, would affect the precision of parameter estimation. That is to say, we need to correct the frequency walk when $|a_2|NT > P/(NT)$ holds. Then a new Doppler KT is proposed to correct the frequency walk. Since the conventional range KT is implemented in the range-frequency and azimuth-time domain, the Doppler KT should be realized in the intra-segment time and inter-segment time domain accordingly. Equation (10) can be rewritten as

$$x(t_q', t_p) \approx \sigma_2 \exp\left\{j2\pi\left[a_0 + a_1(t_q' + NT + t_p) + a_2\left[2(t_q' + NT)t_p + t_p^2\right]/2\right]\right\} \quad (12)$$

where $t_q' = t_q - NT$.

Substituting the scaling expression $(t_q' + NT)t_p = NTt_p'$ into (12) yields

$$x'(t_q', t_p') = \sigma_2 \exp(j2\pi a_0)\exp\left[j2\pi a_1(t_q' + NT)\right]\exp\left(j2\pi a_1 \frac{NT}{t_q' + NT} t_p'\right)$$
$$\times \exp(j\pi 2a_2 NT t_p')\exp\left\{j\pi a_2 \left[\frac{NT}{t_q' + NT} t_p'\right]^2\right\} \quad (13)$$

Since $t_q' \ll NT$, (13) can be further expressed as

$$x'(t_q', t_p') = \sigma_2 \exp(j2\pi a_0)\exp\left[j2\pi a_1(t_q' + NT)\right]\exp(j2\pi a_1 t_p')$$
$$\times \exp(j\pi 2a_2 NT t_p')\exp(j\pi a_2 t_p'^2) \quad (14)$$

The FFT of $x'(t_q', t_p')$ with $t_q'$ is computed to be

$$x'(f_q', t_p') = \sigma_2 \frac{\sin\left[N\pi(a_1 - f_q')/P\right]}{\sin\left[\pi(a_1 - f_q')\right]} \exp\left[j\pi\left(\frac{N}{P}-1\right)(a_1 - f_q')\right]\exp(j2\pi a_0)$$
$$\times \exp(j2\pi a_1 NT)\exp\left[j2\pi(a_1 + a_2 NT)t_p'\right]\exp(j\pi a_2 t_p'^2) \quad (15)$$

From (15), it can be seen that the frequency walk is corrected completely and the energy of target has been concentrated into the frequency cell whose frequency satisfies $f'_q = a_1$. The azimuth signal remains an LFM signal with the frequency $a_1 + a_2 NT$ and the chirp rate $a_2$. Then applying LVT on $x'(f'_q, t'_p)$ with respect to $t'_p$ yields

$$\left( \hat{f}, \hat{\gamma}, \hat{f}'_q \right) = \arg\max_{f, \gamma, f'_q} \left\{ \mathop{\mathrm{LVT}}_{t'_p} \left[ x'(f'_q, t'_p) \right] \right\} \tag{16}$$

Hence the parameters can be estimated by

$$\begin{cases} \hat{a}_2 = \hat{\gamma} \\ \hat{a}_1 = \hat{f} - \hat{a}_2 NT \end{cases} \tag{17}$$

where $\hat{f}'_q$ is the coarse estimate of $a_1$, which satisfies $\hat{f}'_q \in \left[ \frac{IP}{NT} - \frac{P}{2NT}, \frac{IP}{NT} + \frac{P}{2NT} \right]$, $I \in \left[ -\frac{N}{2P} : 1 : \frac{N}{2P} - 1 \right]$, and $\hat{f}$ is the refined estimate of $\hat{a}_1 + \hat{a}_2 NT$. It should be noted that the estimated frequency $\hat{f}$ and chirp rate $\hat{\gamma}$ satisfy $\hat{f} \in [-P/(4NT), P/(4NT)]$ and $\hat{\gamma} \in [-P/(2NT), P/(2NT)]$, respectively. Generally, the available chirp rate $a_2$ is about $[-P/(2NT), P/(2NT)]$, however, the available frequency $a_1 + a_2 NT$ may be within the range of $[P/(4NT), 1/(2T)]$ or $[-1/(2T), -P/(4NT)]$. Then a modified method is proposed to estimate the parameters of targets precisely.

According to the estimated $\hat{f}'_q$ and $\hat{a}_2$, we can calculate the coarse estimate of $\hat{a}_1 + \hat{a}_2 NT$, which satisfies $\hat{f}'_q + \hat{a}_2 NT \in \left[ \frac{AP}{NT} - \frac{P}{2NT}, \frac{AP}{NT} + \frac{P}{2NT} \right]$, $A \in \left[ -\frac{N}{2P} : 1 : \frac{N}{2P} - 1 \right]$ and then construct the searching frequency function $f_{search} = \hat{f} + k_{amb\_in} P/(2NT)$, $k_{amb\_in} = 2A-1:1:2A+1$ within the range of $\hat{f}'_q + \hat{a}_2 NT$ . The corresponding searching velocity is computed to be $v_{search} = \lambda(f_{search} - \hat{a}_2 NT)/2$. However, in the real situation, $k_{amb\_in}$ is selected to be $k_{amb\_in} = 2(A-1)-1:1:2(A+1)+1$ to ensure the correctness of parameter estimates. Then constructing the phase-compensated function

$$H_{com}(t,f) = \exp\left[-j4\pi(f+f_c)\left(v_{search}t + \hat{a}_2\lambda t^2/4\right)/c\right]$$ with $v_{search}$ and multiplying it by (5) yields

$$\begin{aligned} S_{com}(t,f) &= S(t,f)H_{com}(t,f) \\ &= \frac{\sigma Gw(t)}{B}\text{rect}\left(\frac{f}{B}\right)\exp\left\{-j4\pi\frac{f+f_c}{c}\left[R_B - (v_c - v_{search})t\right]\right\} \end{aligned} \quad (18)$$

Hence $v_c$ can be determined by solving

$$v_c = \arg\max_{v_{search}}\left\{\text{sum}_t\left\{\text{IFFT}_f\left[S_{com}(t,f)\right]\right\}\right\} \quad (19)$$

TABLE I SYSTEM PARAMETERS FOR SIMULATION

| System Parameters | Values |
| --- | --- |
| Wavelength (GHz) | 10 |
| Range bandwidth (MHz) | 8 |
| Pulse repetition frequency (Hz) | 2000 |
| Range sampling frequency (MHz) | 20 |
| Pulse width (us) | 20 |

In the following, the effectiveness of the proposed method is examined under the ideal circumstance. The simulation parameters are listed in Table I. The relative radial velocity and acceleration between the slow moving target and the radar platform are $v_c = 10\,\text{m/s}$ and $a_c = 0.92\,\text{m/s}^2$, respectively. Fig. 2(a) shows the trajectory of the target after range compression. It is obvious that the signal energy spreads over several range cells. We perform the SKT to correct the RCM and obtain the result in Fig. 2(b). It is observed that the RCM is eliminated completely. Then segmental FFT is applied to the azimuth signal with the number of segments of 256 and the frequency walk occurs shown in Fig. 2(c). The detailed criterion to choose the number of segments is discussed in Section IV-B. The Doppler KT is used to correct the frequency walk and the result is shown in Fig. 2(d), from which it can be noted that the frequency walk is removed completely. After LVT, as shown in Fig. 2(e), the target is well focused. The frequency and chirp rate with the value of $-19.32\,\text{Hz}$ and $61.55\,\text{Hz/s}$, respectively, are also estimated. Fig. 2(f) shows the searching result of the inner ambiguity number within the range from $-PRF/2$ to $PRF/2$, in which the inner ambiguity number can be easily determined with the value of $k_{amb\_in}=14$. According to the aforementioned analysis, the final estimates of the relative radial velocity and acceleration between the slow moving target and the radar platform are $9.9991\,\text{m/s}$ and $0.9232\,\text{m/s}^2$, respectively.

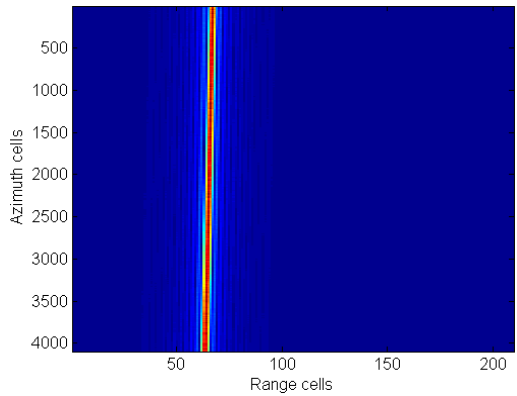 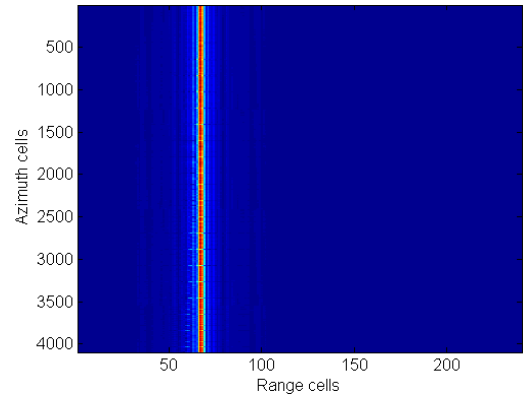

(a)                                          (b)

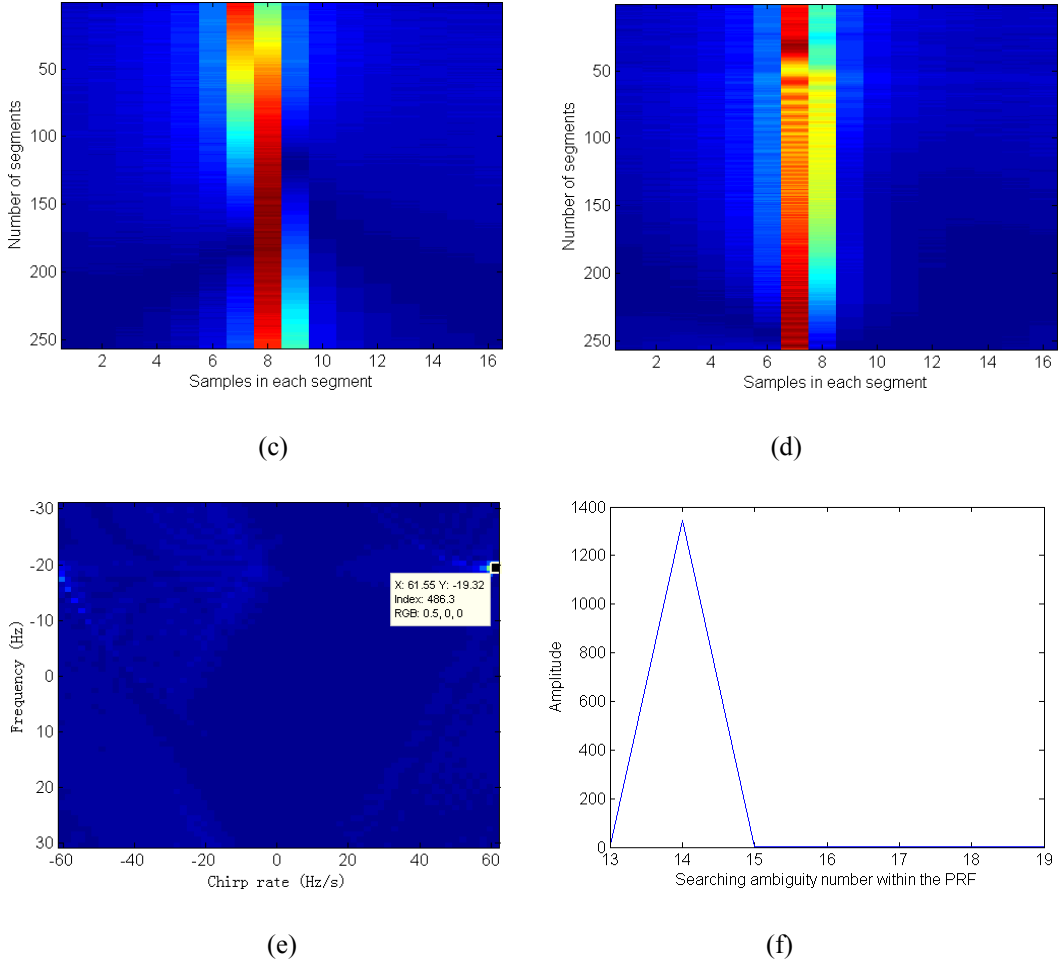

Fig. 2. Simulation results of the slow moving target. (a) Trajectory after range compression. (b) Trajectory after SKT. (c) Trajectory after FFT applied on the azimuth signal within each segment. (d) Trajectory after Doppler KT. (e) Result of LVT. (f) Estimation of inner ambiguity number $k_{amb\_in}$ within the range from $-PRF/2$ to $PRF/2$.

*B. Parameter Estimation Method for the Fast Moving Target*

For a fast moving target, its Doppler frequency will exceed the mission PRF. In this case, the target spectrum will be overlapped by the mission PRF. The fast moving target satisfies $v \in k_{amb\_out} PRF\lambda/2 + [-PRF\lambda/4, PRF\lambda/4]$, where $k_{amb\_out} \neq 0$ denotes the ambiguity number relative to PRF. In this situation, the aforementioned SKT cannot deal with the RCM completely. The velocity of target can be written as

$$v = k_{amb\_out} v_{amb} + v_0 \qquad (20)$$

where $v_{amb} = \text{PRF}\lambda/2$ is the blind velocity and $v_0 \in [-v_{amb}/2, v_{amb}/2]$. Applying SKT to correct the linear RCM and removing the quadratic RCM in the azimuth frequency domain, we can get

$$\begin{aligned}s(t_{a_{seg}}, \tau) = &\sigma G w(t_{a_{seg}}) \text{sinc}\left[\pi B\left(\tau - 2\frac{R_B - k_{a\_out} v_{amb} t_{a_{seg}}}{c}\right)\right] \\ &\times \exp\left[-j4\pi \frac{R_B - v_c t_{a_{seg}} + (V - v_a)^2 t_{a_{seg}}^2/(2R_B)}{\lambda}\right]\end{aligned} \qquad (21)$$

It can be noted that the trajectory in the range-time and azimuth-time domain exhibits linear feature and its slope is proportional to the ambiguity number. Therefore, the RCMC/integration method can be well adopted to estimate the slope [25]. This estimator is formulated as

$$f(k_{amb\_out}, \tau) = \sum_{seg=1}^{P} \sum_{t_{a_{seg}}=(seg-1)\frac{NT}{P}}^{(seg-1)\frac{NT}{P}+(\frac{N}{P}-1)T} \left|\text{IFFT}_f\left[\text{FFT}_\tau\left[s(t_{a_{seg}}, \tau)\right] \exp\left(-j\frac{4\pi f}{c} k_{amb\_out} v_{amb} t_{a_{seg}}\right)\right]\right|$$

$$(22)$$

where IFFT denotes the inverse fast Fourier transform. Then the entropy of an image is employed to determine the estimated value and evaluate the estimation performance. The ambiguity number can be estimated by

$$\begin{cases} E(k_{amb\_out}) = -\sum_\tau \left\{\left[\frac{|f(k_{amb\_out}, \tau)|^2}{\sum_\tau |f(k_{amb\_out}, \tau)|^2}\right] \log\left[\frac{|f(k_{amb\_out}, \tau)|^2}{\sum_\tau |f(k_{amb\_out}, \tau)|^2}\right]\right\} \\ k_{a\_out} = \arg\max_{k_{amb\_out}} \left(\frac{1}{E(k_{amb\_out})}\right) \end{cases} \qquad (23)$$

What should be pointed out is that the ambiguity number can be estimated accurately and the computational load is relatively low because the number value is an integer.

By performing the entropy of an image, we can get a reliable result of ambiguity number in medium- to high- SNR scenarios, however, we cannot obtain the right estimate in low SNR scenario. Accordingly, an improved method is proposed to estimate the ambiguity number.

The phase-compensated function is first constructed as

$$H_{com2}(k_{amb\_out}, t_{a_{seg}}) = \exp\left(-j\frac{4\pi f}{c} k_{amb\_out} v_{amb} t_{a_{seg}}\right) \quad (24)$$

Multiplying (24) by the signal after RCM correction in the range-frequency and azimuth-time domain yields

$$S'(t_{a_{seg}}, f) \approx \frac{\sigma G w(t_{a_{seg}})}{B} \text{rect}\left(\frac{f}{B}\right) \exp\left[-j\frac{4\pi f}{c}\left(k_{amb\_out} - k_{a\_out}\right) v_{amb} t_{a_{seg}}\right] \\ \times \exp\left\{-j4\pi \frac{f+f_c}{c} R_B + j4\pi \frac{f_c}{c} v_c t_{a_{seg}} - j2\pi \frac{f_c}{c} \frac{(V-v_a)^2}{R_B} t_{a_{seg}}^2 \right\} \quad (25)$$

Applying the IFFT on $S'(t_{a_{seg}}, f)$ with respect to $f$ yields

$$s(t_{a_{seg}}, \tau) = \sigma G w(t_{a_{seg}}) \text{sinc}\left[\pi B \left(\tau - 2\frac{R_B - (k_{a\_out} - k_{amb\_out}) v_{amb} t_{a_{seg}}}{c}\right)\right] \\ \times \exp\left[-j4\pi \frac{R_B - v_c t_{a_{seg}} + (V-v_a)^2 t_{a_{seg}}^2/(2R_B)}{\lambda}\right] \quad (26)$$

Then the Doppler LVT is applied on (26) with respect to $t_{a_{seg}}$ and the ambiguity number $k_{a\_out}$ is estimated as

$$\left(k_{a\_out}, \hat{f}, \hat{\gamma}, \hat{f}_q'\right) = \underset{k_{amb\_out}, f, \gamma, f_q'}{\arg\max} \underset{t_p'}{\text{LVT}}\left[x'(f_q', t_p', k_{amb\_out})\right] \quad (27)$$

where $x'(f_q', t_p', k_{amb\_out})$ is the derived azimuth signal of $s(t_{a_{seg}}, \tau)$ corresponding to different ambiguity number $k_{amb\_out}$. The derivation of $x'(f_q', t_p', k_{amb\_out})$ is the same as that in Section III-A.

It can be seen from (27) that the accurate parameter estimates can be obtained if the parameter $k_{amb\_out}$ is matched with the ambiguity number $k_{a\_out}$. Otherwise, the smearing result will be obtained. The parameters can be further estimated by

$$\begin{cases} \hat{a}_2 = \hat{\gamma} \\ \hat{a}_1 = \hat{f} - \hat{a}_2 NT \end{cases} \tag{28}$$

where $\hat{f}'_q$ is the coarse estimate of $a_1$ within the range from $-1/(2T)$ to $1/(2T)$, which satisfies $\hat{f}'_q \in \left[ \frac{IP}{NT} - \frac{P}{2NT}, \frac{IP}{NT} + \frac{P}{2NT} \right]$, $I \in \left[ -\frac{N}{2P} : 1 : \frac{N}{2P} - 1 \right]$, and $\hat{f}$ is the refined estimate of $\hat{a}_1 + \hat{a}_2 NT$. It should be noted that the equivalent interval where $\hat{a}_1 + \hat{a}_2 NT$ is located in turns into $[k_{amb\_out}/T - P/(4NT), k_{amb\_out}/T + P/(4NT)]$, however, the available frequency $a_1 + a_2 NT$ may be within the range of $[k_{amb\_out}/T + P/(4NT), k_{amb\_out}/T + 1/(2T)]$ or $[k_{amb\_out}/T - 1/(2T), k_{amb\_out}/T - P/(4NT)]$.

According to the estimated $\hat{f}'_q$ and $\hat{a}_2$, we can calculate the coarse estimate of $\hat{a}_1 + \hat{a}_2 NT$, which satisfies

$$\frac{k_{amb\_out}}{T} + \hat{f}'_q + \hat{a}_2 NT \in \left[ \frac{k_{amb\_out}}{T} + \frac{AP}{NT} - \frac{P}{2NT}, \frac{k_{amb\_out}}{T} + \frac{AP}{NT} + \frac{P}{2NT} \right], \quad A \in \left[ -\frac{N}{2P} : 1 : \frac{N}{2P} - 1 \right]$$

and then construct the searching frequency function $f_{search} = \hat{f} + k_{amb\_in} P/(2NT)$, $k_{amb\_in} = 2Nk_{amb\_out}/P + 2A - 1 : 1 : 2Nk_{amb\_out}/P + 2A + 1$ within the range of $\hat{f}'_q + \hat{a}_2 NT$. The corresponding searching velocity is computed to be $v_{search} = \lambda(f_{search} - \hat{a}_2 NT)/2$. However, in the real situation, $k_{amb\_in}$ is selected to be $k_{amb\_in} = 2Nk_{amb\_out}/P + 2(A-1) - 1 : 1 : 2Nk_{amb\_out}/P + 2(A+1) + 1$ to ensure the correctness of parameter estimates. The subsequent steps are the same as that in Section III-A.

Fig. 3 shows the result of the proposed method for the fast moving target with the relative radial velocity of $v_c = 40 \, \text{m/s}$ and the relative radial acceleration of $a_c = 0.92\,\text{m/s}^2$. Fig. 3(a) shows the trajectory of the target after range compression. It can be seen that the signal energy spreads over a large number of range cells during the exposure time. Fig. 3(b) shows the result after RCM correction is performed. It is observed that the RCM cannot be well mitigated because of the ambiguous velocity. Fig. 3(c) shows the reciprocal of the entropy of the RCMC/integration. The ambiguity number can be easily determined with the right value of $k_{a\_out} = 1$. Fig. 3(d) shows the result of RCM correction after the phase compensation with the estimated ambiguity number, from which it can be seen that the large RCM is eliminated completely. Then segmental FFT is applied on the azimuth signal with the number of segments of 256 and the frequency walk occurs shown in Fig. 3(e). The Doppler KT is used to correct the frequency walk and the result is shown in Fig. 3(f), from which it can be noted that the frequency walk is removed completely. After LVT, as shown in Fig. 3(g), the target is well focused. The frequency and chirp rate with the value of $-19.32\,\text{Hz}$ and $61.55\,\text{Hz/s}$, respectively, are also estimated. Fig. 3(h) shows the searching result of the inner ambiguity number within the frequency range centered at $PRF$ from $PRF/2$ to $3PRF/2$, in which the inner ambiguity number can be easily determined with the value of $k_{amb\_in} = 46$. According to the aforementioned analysis, the final estimates of the relative radial velocity and acceleration between the fast moving target and the radar platform are $39.9991\,\text{m/s}$ and $0.9232\,\text{m/s}^2$, respectively.

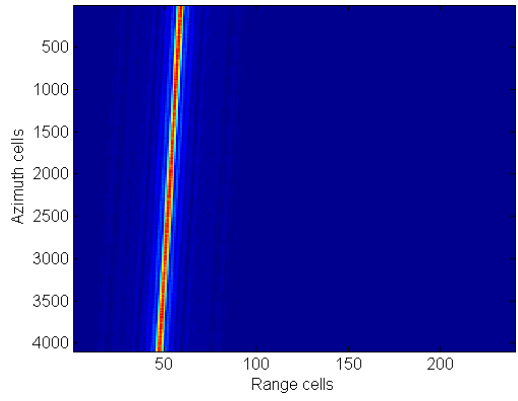 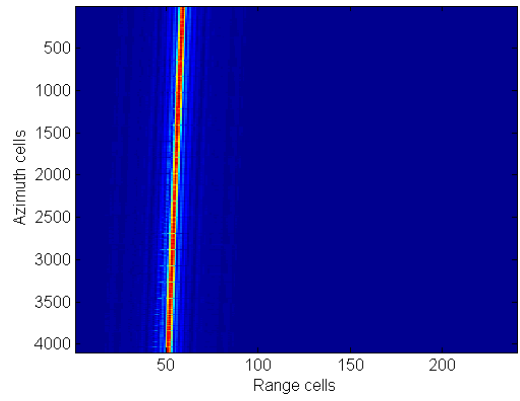

(a)                          (b)

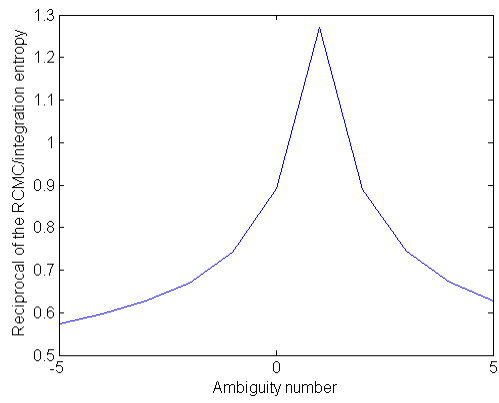 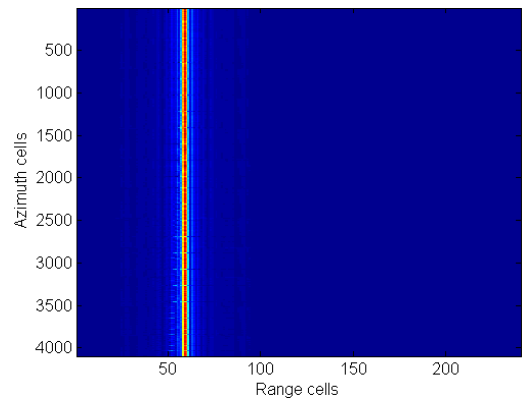

(c)                          (d)

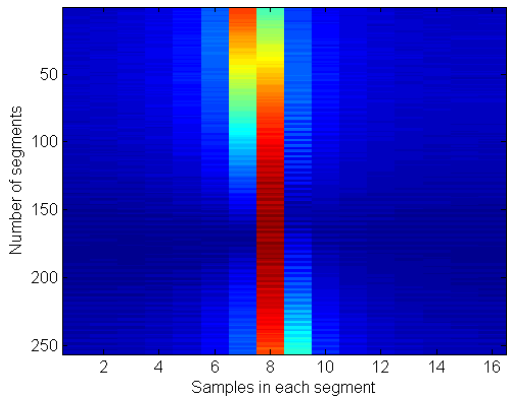 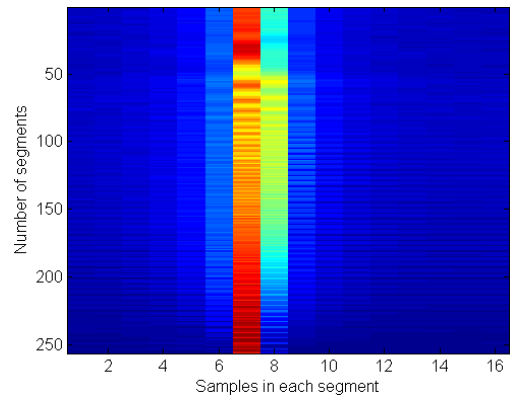

(e)                          (f)

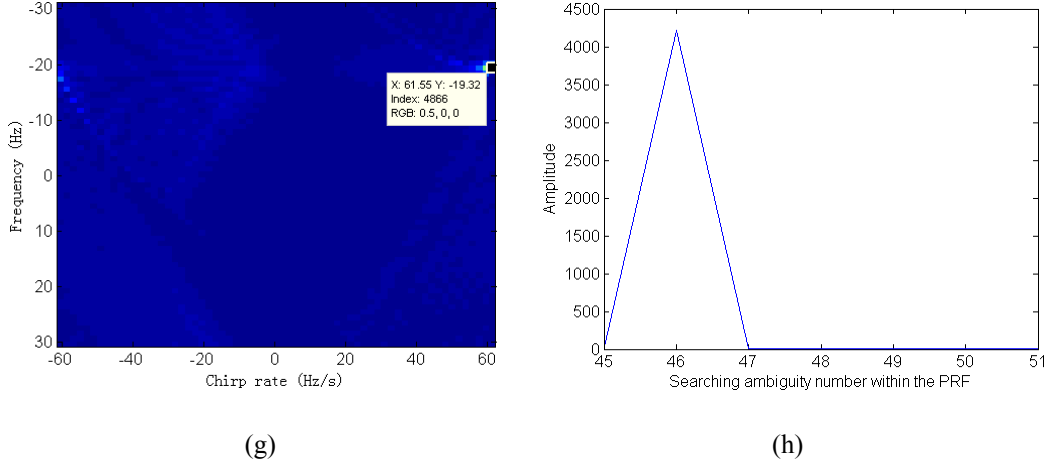

(g)                  (h)

Fig. 3. Simulation results of the fast moving target. (a) Trajectory after range compression. (b) Trajectory after RCM correction. (c) Estimation of the ambiguity number $k_{a\_out}$. (d) Trajectory after RCM correction and ambiguity number compensation. (e) Trajectory after FFT applied on the azimuth signal within each segment. (f) Trajectory after Doppler KT. (g) Result of LVT. (h) Estimation of the inner ambiguity number $k_{amb\_in}$ within the range from $PRF/2$ to $3PRF/2$.

## IV APPLICATIONS AND DISCUSSIONS

### A. Implementation of the SKT

It is worth to mention that the conventional KT aligns the peak position of the echo envelope in each pulse repetition time (PRT) of each segment to that in the first PRT of that segment [16, 31]. We present the KT processing in each segment during the exposure time in the following.

After rang compression, the received signal of the $seg$-th segment in the range-frequency and azimuth-time domain can be expressed as

$$S(t_{seg}, f) = \frac{\sigma G w(t_{seg})}{B} \mathrm{rect}\left(\frac{f}{B}\right) \\ \times \exp\left\{-j4\pi \frac{f+f_c}{c}\left[R_B - v_c t_{seg} + (V-v_a)^2 t_{seg}^2 \big/(2R_B)\right]\right\} \quad (29)$$

where $t_{seg} \in [(seg-1)NT/P : (seg-1)NT/P + (N/P-1)T]$.

Let $t_{seg}$ be $t_{seg} = t_{in} + (seg-1) \cdot NT/P$ with $t_{in} = (0 : N/P-1)T$ and $seg = 1, 2, ..., P$. Then (29) is further expressed as

$$S(t_{in}, seg, f) = \frac{\sigma Gw[t_{in} + (seg-1)NT/P]}{B} \text{rect}\left(\frac{f}{B}\right)$$

$$\times \exp\left\{-j4\pi \frac{f+f_c}{c}\left[R_B - v_c\left[t_{in} + (seg-1)\frac{NT}{P}\right] + \frac{(V-v_a)^2[t_{in} + (seg-1)NT/P]^2}{2R_B}\right]\right\}$$

$$= \frac{\sigma Gw[t_{in} + (seg-1)NT/P]}{B} \text{rect}\left(\frac{f}{B}\right)$$

$$\times \exp\left\{-j4\pi \frac{f+f_c}{c}\left[R_B - v_c(seg-1)\frac{NT}{P} + \frac{(V-v_a)^2[(seg-1)NT/P]^2}{2R_B}\right]\right\}$$

$$\times \exp\left\{-j4\pi \frac{f+f_c}{c}\left[R_B - v_c t_{in} + (V-v_a)^2 \frac{t_{in}^2 + 2t_{in}(seg-1)NT/P}{2R_B}\right]\right\}$$

(30)

Substituting the scaling factor $t_{in} = \frac{f_c}{f+f_c} t'_{in}$ into the signal of the $seg$-th segment yields

$$S(t'_{in}, seg, f)$$
$$= \frac{\sigma Gw[t_{in} + (seg-1)NT/P]}{B}\text{rect}\left(\frac{f}{B}\right)$$
$$\times \exp\left\{-j4\pi \frac{f+f_c}{c}\left[R_B - v_c(seg-1)NT/P + (V-v_a)^2((seg-1)NT/P)^2/(2R_B)\right]\right\} \quad (31)$$
$$\times \exp\left[j\frac{4\pi}{\lambda}\left(v_c - \frac{(V-v_a)^2 NT}{R_B P}(seg-1)\right)t'_{in}\right]\exp\left[-j\frac{2\pi}{\lambda}\frac{(V-v_a)^2}{R_B}\left(1-\frac{f}{f_c}\right)t'^{2}_{in}\right]$$

Take the Sinc interpolation to realize KT and we have

$$S(m', seg, f) = \sum_{m=1}^{N/P} S(m, seg, f)\text{sinc}\left(\frac{f_c}{f+f_c}m' - m\right) \quad (32)$$

Performing IFFT on $S(t'_{in}, seg, f)$ with $f$ yields

$$s(t'_{a_{seg}}, seg, \tau) =$$

$$\sigma Gw(t'_{a_{seg}})\mathrm{sinc}\left[\pi B\left(\tau - 2\frac{R_{seg} - (V-v_a)^2\left(t'_{a_{seg}} - (seg-1)NT/P\right)^2 / (2R_B)}{c}\right)\right] \quad (33)$$

$$\times \exp\left[-j4\pi\frac{R_B - v_c t'_{a_{seg}} + (V-v_a)^2 t'^2_{a_{seg}}/(2R_B)}{\lambda}\right]$$

where $t'_{a_{seg}} = t'_{in} + (seg-1)NT/P$ and

$R_{seg} = R_B - v_c(seg-1)NT/P + (V-v_a)^2\left[(seg-1)NT/P\right]^2/(2R_B)$.

It can be seen from (33) that after SKT operation, the peak position of the envelope of different segment is aligned to different rang cells, which degrades the performance of the proposed method. An intuitive method to deal with this problem is making the position of alignment of KT processing controllable, thereby migrating the peak position of the envelope of different segment to the same range cell. Hence, a modified realization of SKT is proposed to ensure (6) holds. The SKT implementation using Sinc interpolation is modified as

$$S(m', seg, f) = \sum_{m=1}^{N/P} S(m, seg, f)\mathrm{sinc}\left[\frac{f_c}{f+f_c}\left(m' + (seg-1)\frac{N}{P}\right) - \left(m + (seg-1)\frac{N}{P}\right)\right] \quad (34)$$

By using (34), the peak position of the envelope in each PRT of each segment is aligned to that in the first PRT of the first segment. Since KT is essentially a uniform resampling or a linear scaling, it can be carried out efficiently by chirp transform [24], Chirp-Z transform [32] or scaled fast FT [33]. In these transforms, the scaling factor is updated with the range frequency. These implementations are interpolation free and use only complex multiplications. Fig. 4 shows the flowchart of the SKT through Chirp-Z transform.

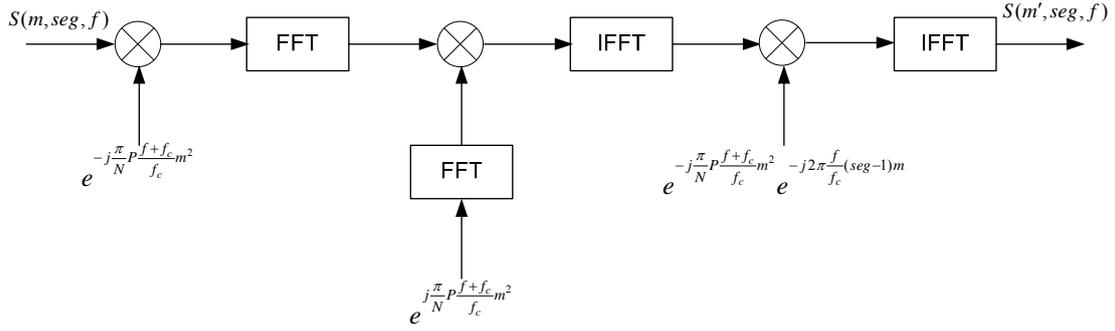

Fig. 4. The flowchart of the SKT through Chirp-Z transform.

*B. the Criterion to Choose the Number of Segments*

As discussed above, we need to do segment processing of the azimuth signal to estimate the parameters. And the choice of the number of segments would affect the integration gain of each segment and the accuracy of the parameter estimation. The criterion of deciding the number of segments is given as follows.

According to the assumption of the derivation, i.e., neglecting the change of frequency during the interval of each segment, and to achieve the integration gain with the lowest integration loss in each segment, the integration interval $NT/P$ of each segment should be less than $|1/\Delta f(NT/P)|$ with $\Delta f(NT/P)$ denoting the change of frequency within the interval of $NT/P$. Therefore, we obtain the selection criterion

$$NT/P < 1/\sqrt{|a_2|} \tag{35}$$

Generally speaking, the choice of *P* is a tradeoff among the acceptable performance degradation, the tolerable computational complexity and memory cost, under the condition of satisfying (35).

*C. Processing for Moving Targets with Marginal Velocity*

The spectrum of the moving targets with marginal velocity [9] is split into two parts by the mission PRF and spans the neighboring PRF bands. After range migration correction, the signal in the range-time and azimuth-time domain can be represented as

$$\begin{cases} s(t_{a_{seg}},\tau) = s_1(t_{a_{seg}},\tau) + s_2(t_{a_{seg}},\tau) \\ s_1(t_{a_{seg}},\tau) = \sigma G_1 w(t_{a_{seg}}) \text{sinc}\left[\pi B\left(\tau - 2\frac{R_B - (k_{a\_out}-1)v_{amb}t_{a_{seg}}}{c}\right)\right] \\ \qquad \times \exp\left[-j4\pi \frac{R_B - v_c t_{a_{seg}} + (V-v_a)^2 t_{a_{seg}}^2/(2R_B)}{\lambda}\right] \\ s_2(t_{a_{seg}},\tau) = \sigma G_2 w(t_{a_{seg}}) \text{sinc}\left[\pi B\left(\tau - 2\frac{R_B - k_{a\_out}v_{amb}t_{a_{seg}}}{c}\right)\right] \\ \qquad \times \exp\left[-j4\pi \frac{R_B - v_c t_{a_{seg}} + (V-v_a)^2 t_{a_{seg}}^2/(2R_B)}{\lambda}\right] \end{cases} \quad (36)$$

where $G_1$ and $G_2$ are the gain of the range compression for the two parts [located at the $(k_{a\_out}-1)$th and $k_{a\_out}$th PRF], respectively.

From (36), it is evident that, two straight lines exist with different slopes expressed as

$$\begin{cases} \mu_1 = 2\frac{(k_{a\_out}-1)v_{amb}}{c} \\ \mu_2 = 2\frac{k_{a\_out}v_{amb}}{c} \end{cases} \quad (37)$$

If $k_{amb\_out}$ in the constructed compensation function satisfies $k_{amb\_out} = k_{a\_out} - 1$, the signal energy of $s_1(t_{a_{seg}},\tau)$ can be accumulated effectively and the correct parameter estimates can be obtained, while the signal energy of $s_2(t_{a_{seg}},\tau)$ cannot be accumulated completely, resulting in the defocused target. In the same way, if $k_{amb\_out}$ in the constructed compensation function satisfies $k_{amb\_out} = k_{a\_out}$, the signal energy of $s_2(t_{a_{seg}},\tau)$ can be accumulated effectively and the correct parameter estimates can be obtained, while $s_1(t_{a_{seg}},\tau)$ will be defocused.

That is, although the signal energy of different part can be accumulated individually, the energy of each part is less than the total energy. This phenomenon is disadvantageous to the parameter estimation. To avoid these deficiencies, preprocessing should be performed on the signal of target before estimating parameters in this case. For the target with the Doppler bandwidth smaller than $1/(2T)$, Doppler shifting by $1/(2T)$ is implemented to ensure the spectrum of the signal is not split into two parts. The accurate implementation consists of the following major steps. First, the compensation function is constructed as

$$H_{com3}(t,f) = \exp\left[-j\frac{\pi(f+f_c)}{Tf_c}t\right] \tag{38}$$

And then multiplying (38) by (5) yields

$$S(t,f) = \frac{\sigma G w(t)}{B}\text{rect}\left(\frac{f}{B}\right)\exp\left\{-j4\pi\frac{f+f_c}{c}\left[R_B - \left(v_c - \frac{\lambda}{4T}\right)t + (V-v_a)^2 t^2/(2R_B)\right]\right\} \tag{39}$$

It is observed that the target spectrum becomes an entire part. After that, the proposed method in Section III is applied on (39) to achieve the estimates of the parameters. The simulated data is employed to examine the correctness in this general case. Fig. 5 shows the results of a moving target with marginal velocity. The result after SKT operation is shown in Fig. 5(a), from which we can find two trajectories with different slopes. Fig. 5(b) shows the signal after azimuth spectrum compression, from which it can be seen that the spectrum is split into two parts. Fig. 5(c) shows the trajectory of the target after Doppler shifting. From this figure, it can be seen that the trajectory turns into a straight line. Fig. 5(d) shows the result of azimuth spectrum compression applied on the signal after Doppler shifting. It can be seen that the Doppler spectrum is not split into two parts after shifting.

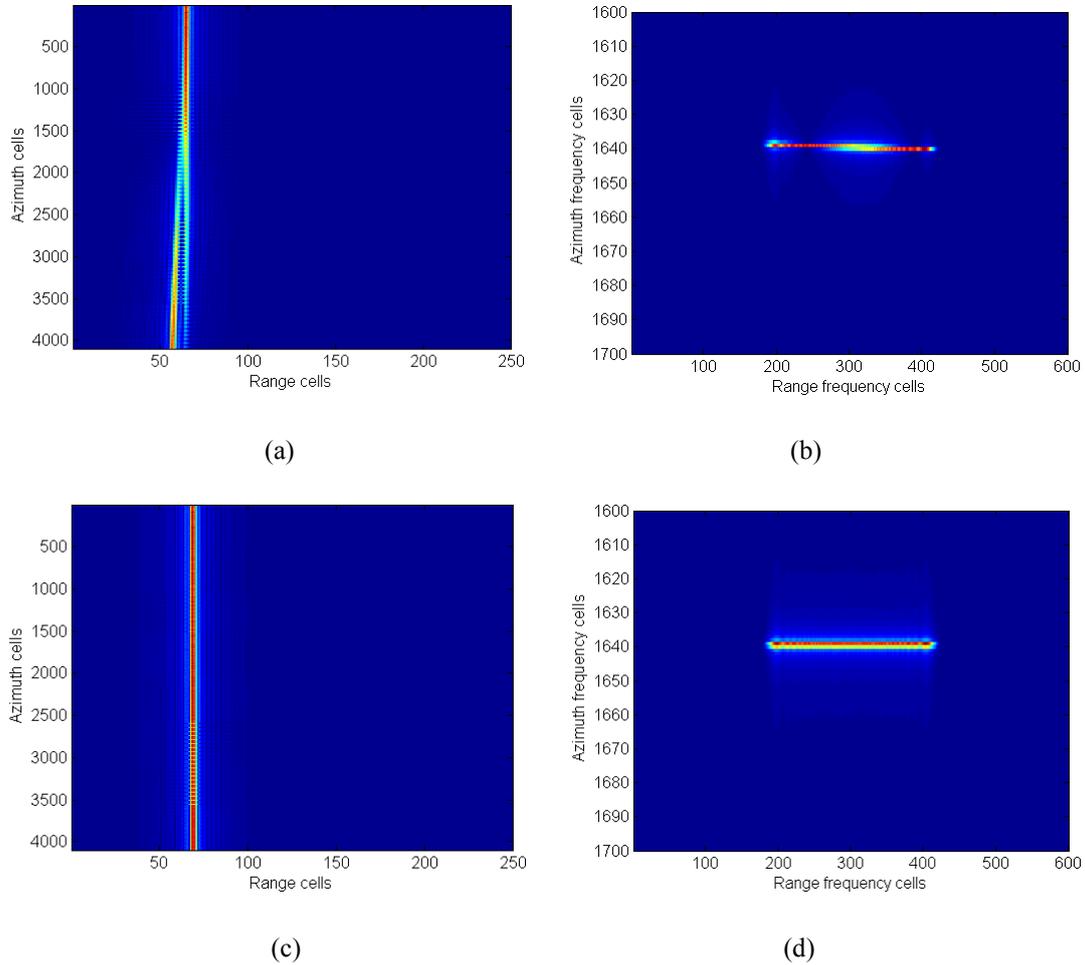

Fig.5. Simulation results of a moving target with margin velocity. (a) Trajectory after SKT. (b) Compressed azimuth spectrum. (c) Trajectory after SKT and Doppler shifting. (d) Compressed azimuth spectrum after Doppler shifting.

*D. Processing for Multiple Moving Targets*

From the aforementioned analysis, it is known that the proposed method can directly focus a slow moving target without knowing its motion parameters; while for a fast moving target, we just need to know its ambiguity number. For multiple moving targets with the same ambiguity number, phase compensation function is constructed with (24) and the precise parameter estimates can be achieved simultaneously. While for multiple moving targets with different ambiguity number, the phase compensation factors should be constructed respectively. In this way, a moving target is expected to be well focused after compensating the phase related with the ambiguity number and to be defocused by a mismatched factor. The mismatching of (24) will result in a residual linear RCM and thus introduce defocusing. In this case, the different constructed phase compensation function $H_{com2}(k_{amb\_out}, t_{a_{seg}})$ and $H_{com}(t, f)$ are employed to achieve the parameter estimates of each target. If the scattering intensities of multiple targets differ significantly, the clean technique [34] is employed to improve the precision of the estimates.

To investigate the effectiveness of the proposed method, three targets are set to be located in the same range cell. The relative radial velocity and acceleration between the targets and the platform are $10\,\text{m/s}$, $10\,\text{m/s}$, $9\,\text{m/s}$, $0.9\,\text{m/s}^2$, $0.93\,\text{m/s}^2$ and $0.93\,\text{m/s}^2$, respectively. It is seen that the relative radial velocities of target 1 and target 2 and the relative radial accelerations of target 2 and target 3 are, respectively, identical, which are selected to better explain how the new approach works. Figs. 6(a) and 6(b) are the results of the pulse compression and the proposed method for the three targets, respectively. It is obvious that the three targets can be well focused and the estimates of the three targets are $10.0083\,\text{m/s}$, $10.0206\,\text{m/s}$, $9.0115\,\text{m/s}$, $0.8946\,\text{m/s}^2$, $0.9232\,\text{m/s}^2$ and $0.9232\,\text{m/s}^2$, respectively.

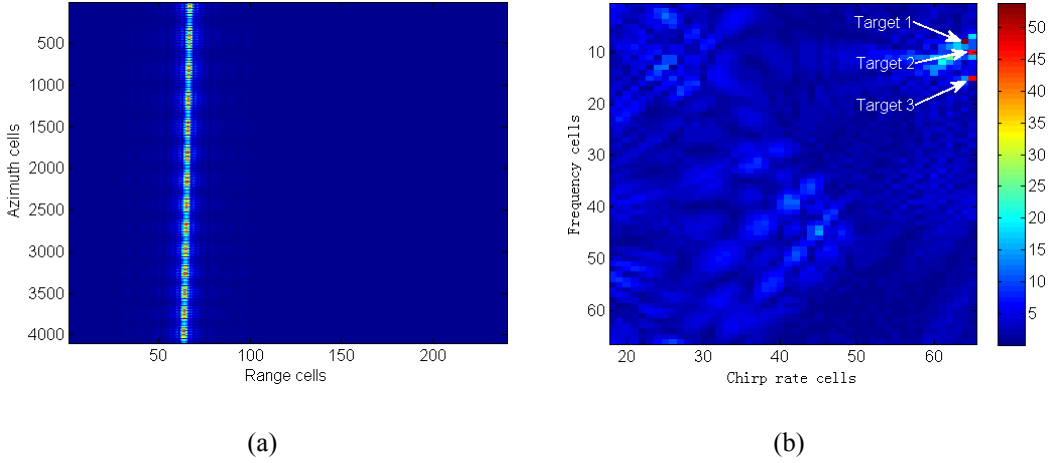

(a)                      (b)

Fig. 6. Results for multiple targets. Result after (a) pulse compression and (b) Doppler LVT.

*E. Computational Complexity and Memory Cost*

To reduce the computational complexity and memory cost, segment processing is introduced for parameter estimation. The SKT and Doppler LVT operation can be implemented through parallel processing to reduce the storage memory cost. In addition, the operation of sliding window could be used to select the data of Doppler LVT processing, which can further reduce the storage memory requirement for parameter estimation over long observation interval. In many practical radar systems, the selection criteria of sliding window can be found in [35].

In what follows, the computational complexity of the proposed Doppler LVT and the direct LVT (DT-LVT) in [20] will be analyzed. As to the intra-segment FFT, $P \frac{N}{2P} \log_2 \left( \frac{N}{P} \right)$ multiplications are needed. For the Doppler KT and the inter-segment LVT operation, $\frac{7}{2} N \log_2 P$ and $2 \frac{N}{P} P^2 \log_2 P$ multiplications are needed, respectively. Therefore, the overall complexity of the Doppler LVT is $N(2P \log_2 P + \frac{7}{2} \log_2 P + \frac{1}{2} \log_2 \frac{N}{P})$. And the complexity of the DT-LVT is $2N^2 \log_2 N$. Defining $\eta$ as the complexity ratio of the Doppler LVT to DT-LVT, the complexity ratio is computed to be $\eta \approx P \log_2 P / (N \log_2 N)$ according to the aforementioned analysis. Taking $N = 4096$ and $P = 256$ for example, the reduced complexity can be $\eta = 4.17\%$, which suggests that the complexity of the new approach is reduced significantly, making this approach more suitable for real-time processing.

*F. Output SNR*

The detection performance can be examined in terms of output SNR; therefore, the output SNR of the proposed method is derived and analyzed. According to [20], the output SNR after DT-LVT operation is limited by

$$SNR_{out1} > \frac{(N\sigma_2)^4 / 4}{(N\sigma_2)^2 N\sigma_N^2 / 2 + (N\sigma_N^2)^2 / 2} = \frac{N^2 SNR_{in}^2 / 2}{NSNR_{in} + 1} \tag{40}$$

where $SNR_{in} = \sigma_2^2 / \sigma_N^2$ is the input SNR of the azimuth signal, and $N$ is the number of pulses during the exposure time.

Next we derive the output SNR of the proposed method. It is indicated in (15) that the energy of the target in each segment has been concentrated into the frequency resolution bin satisfying $f_q' = a_1$ after intra-segment FFT and its output SNR is $SNR_{FFT} = \left(\frac{N}{P}\sigma_2\right)^2 / \left(\frac{N}{P}\sigma_N^2\right)$, where $\left(\frac{N}{P}\sigma_2\right)^2$ and $\frac{N}{P}\sigma_N^2$ denote the power of signal and noise after intra-segment FFT, respectively. After the frequency walk correction and the inter-segment LVT operation, the output SNR is limited by

$$SNR_{out2} > \frac{\left(P\frac{N}{P}\sigma_2\right)^4 / 4}{\left(P\frac{N}{P}\sigma_2\right)^2 P\frac{N}{P}\sigma_N^2 / 2 + \left(P\frac{N}{P}\sigma_N^2\right)^2 / 2} \tag{41}$$

(41) can be further simplified as

$$SNR_{out2} > \frac{(N\sigma_2)^4/4}{(N\sigma_2)^2 N\sigma_N^2/2 + (N\sigma_N^2)^2/2} = \frac{N^2 SNR_{in}^2/2}{NSNR_{in}+1} \quad (42)$$

From (40) and (42), it can be seen that the lower limit of the new approach is equal to that of DT-LVT. It should be noted that the additional SNR loss of segment processing is not considered during the theoretical derivation of $SNR_{out2}$. In practical applications, the scalloping loss exists in intra-segment FFT operation. However, it can be decreased through windowed FFT operation or FFT with zero-padding.

V EXPERIMENTAL RESULTS

In this section, some results with simulated and real data are presented to validate the performance of the proposed algorithm and comparisons are performed between the proposed SKT-DLVT and the method in [20] for the slow and fast moving targets.

*A. Simulated Data*

The parameters used in the simulation are listed in Table I. The signal is embedded in complex white Gaussian noise and the input SNR of the target is $SNR = [-44:2:-30] dB$. For each input SNR value, 500 trials are performed to calculate the root-mean-square errors (RMSE) of the estimates of the target for the SKT-DLVT and the method in [20]. Figs. 7(a) and 7(b) show the RMSE of the velocity and acceleration estimates for the slow moving target with $v_c = 10 \, m/s$ and $a_c = 0.92 \, m/s^2$, respectively. Figs. 8(a) and 8(b) show the RMSE of the velocity and acceleration estimates for the fast moving target with $v_c = 40 \, m/s$ and $a_c = 0.92 \, m/s^2$, respectively.

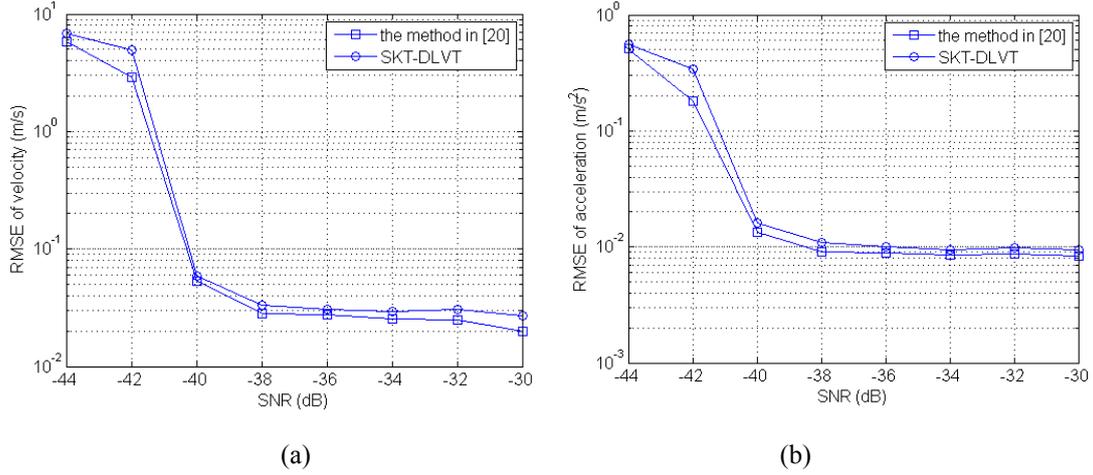

Fig.7. RMSE of (a) velocity and (b) acceleration against input SNRs via the SKT-DLVT and the method in [20] for the slow moving target.

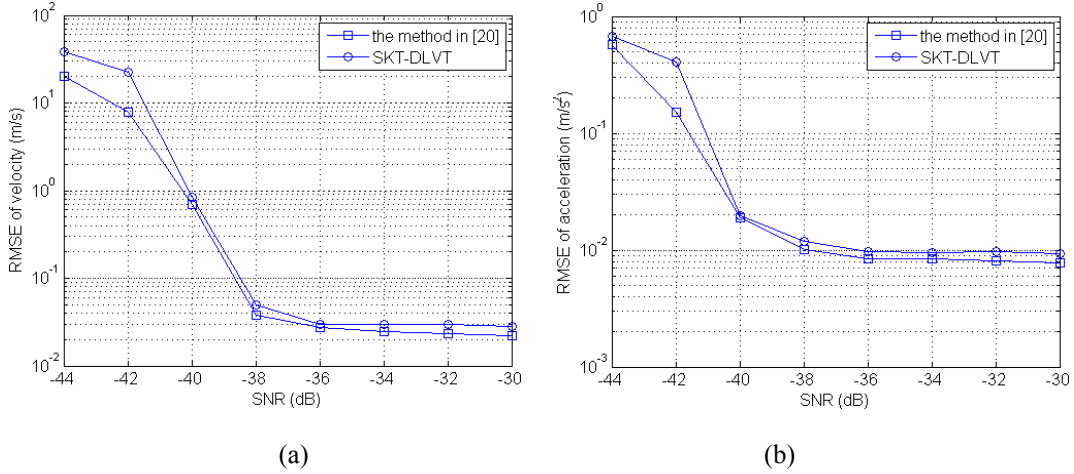

Fig.8. RMSE of (a) velocity and (b) acceleration against input SNRs via the SKT-DLVT and the method in [20] for the fast moving target.

It can be seen from Figs. 7(a), 7(b), 8(a) and 8(b) that the SKT-DLVT has similar performance of parameter estimation to the method in [20]. According to the analysis in Section IV-E, compared with the method in [20], the SKT-FLVT has significantly reduced computational complexity, which makes it feasible for the real-time processing systems.

*B. Real Data*

Part of the RADARSAT-1 Vancouver scene data [3] were selected to verify our proposed method and analysis. The system parameters of these data are given in Table II and the proposed procedure is performed on the selected target (labeled in the Fig. 9(a)). Fig. 9(b) shows the result after SKT, from which it can be seen that the large RCM cannot be eliminated completely because of the velocity ambiguity. Fig. 9(c) shows the reciprocal of the entropy of the RCMC/integration. The ambiguity number can be easily determined with the value of $k_{amb\_out} = -6$. Fig. 9(d) shows the SKT result after the phase compensation with the estimated ambiguity number, from which it can be seen that the large RCM is eliminated completely. After DLVT, as shown in Fig. 9(e), the target is well focused. And the velocity and acceleration with the value of $-204.9606 \, \text{m/s}$ and $50.1447 \, \text{m/s}^2$, respectively, are also estimated. The corresponding frequency and chirp rate are equal to $-7247.0132 \, \text{Hz}$ and $1773.0205 \, \text{Hz/s}$, respectively, which is consistent with the results of the conventional parameter estimation method, thereby verifying the effectiveness of the new approach.

TABLE II SYSTEM PARAMETERS FOR RADARSAT DATA

| System parameters | Values |
|---|---|
| Carrier frequency (GHz) | 5.3 |
| Range bandwidth (MHz) | 30.116 |
| Pulse repetition frequency (Hz) | 1256.98 |
| Range sampling frequency (MHz) | 32.317 |
| Pulse width (us) | 41.74 |
| Doppler centriod frequency (Hz) | -6900 |
| Azimuth chirp rate (Hz/s) | 1733 |

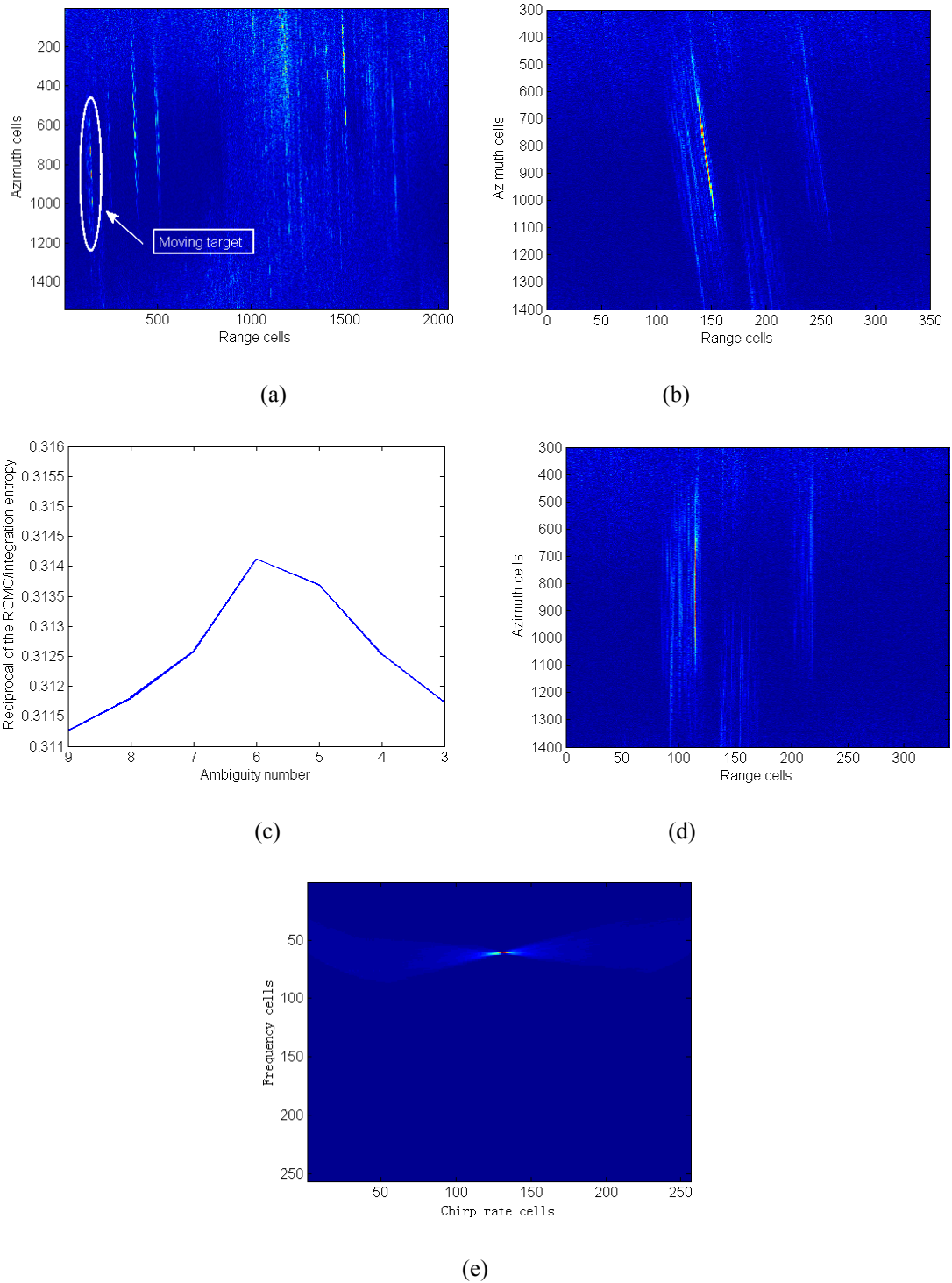

Fig. 9. Result of the Real data via the proposed method. (a) Trajectory after range compression. (b) Trajectory after SKT. (c) Ambiguity number estimation of the target. (d) Trajectory after SKT and ambiguity number compensation. (e) Result of DLVT.

VI C<span>ONCLUSIONS</span>

This paper has introduced a parametric estimation method for the ground moving targets. For the slow moving targets with unambiguous velocity, it can estimate the parameters of targets simultaneously without specific knowledge on the targets' motion. While for the fast moving targets, i.e., in the presence of velocity ambiguity, only its ambiguity number, which can be well estimated by calculating the image entropy (in medium- to high- SNR scenarios) or searching directly within the limited range (in low SNR scenario), is needed, to achieve the parameter estimates precisely. The new approach does not suffer from the considerable troublesome cross-term interference, making it work well for multiple targets. The SKT and DLVT are inherently suitable for parallel implementation and the computations can be parallelized to run on multiple processors with the same (or very similar) program and at the same duration. It can achieve the precise parameter estimation in low SNR scenario because of its effective coherent integration. The performance of the proposed algorithm has been validated by experimental results of simulated data and real data, which shows that the proposed algorithm serves as a good candidate for GMTI. In the near future, algorithms will be designed for parameter estimation of targets with high-order complex motion (i.e., the existence of the along-track acceleration and time-varying cross-track acceleration).

New York: Wiley, 1991.

[3] I. G. Cumming and F. H. Wong, *Digital Processing of Synthetic Aperture Radar Data Algorithms and Implementation*, Norwood, MA: Artech House, 2005.

[4] S. Q. Zhu, G. S. Liao, Y. Qu, Z. G. Zhou, and X. Y. Liu, "Ground moving targets imaging algorithm for synthetic aperture radar," *IEEE Trans. Geosci. Remote Sens.,* vol. 49, no.1, pp. 462-477, Jan. 2011.

[5] C. M. Delphine, K. Jens, R. B. Andreas, and H. G. E. Joachim, "Wide-area traffic monitoring with the SAR/GMTI system PAMIR," *IEEE Trans. Geosci. Remote Sens.,* vol. 46, no. 10, pp. 3019-3030, Oct. 2008.

[6] F. Zhou, R. Wu, M. Xing, and Z. Bao, "Approach for single channel SAR ground moving target imaging and motion parameter estimation," *IET Radar Sonar Navig.,* vol. 1, no. 1, pp. 59-66, Feb. 2007.

[7] G. Sun, M. D. Xing, Y. Wang, F. Zhou, Y. Wu, and Z. Bao, "Improved ambiguity estimation using a modified fractional radon transform," *IET Radar Sonar Navig.,* vol. 5, no. 4, pp. 489-495, Apr. 2011.

[8] G. Li, X. G. Xia and Y. N. Peng, "Doppler keystone transform: an approach suitable for parallel implementation of SAR moving target imaging," *IEEE Geosci. Remote Sens. Lett.,* vol. 5, no. 4, pp. 573-577, Oct. 2008.

[9] G. C. Sun, M. D. Xing, X. G. Xia, Y. R. Wu, and Z. Bao, "Robust ground moving-target imaging using deramp–keystone processing," *IEEE Trans. Geosci. Remote Sens.,* vol. 51, no. 2, pp. 966-982, Feb. 2013.

[10] R. P. Xu, D. D. Zhang, D. H. Hu, X. L. Qiu, and C. B. Ding, "A novel motion parameter
35

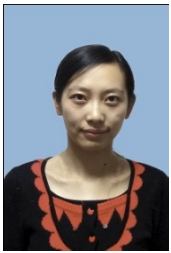
Jing Tian was born in Shandong, China, in November 1984. She received the B.Eng. and M.Sc. degrees both in electronic engineering, from Xidian University, in 2006 and 2009, respectively. She is currently working towards the Ph.D. degree in the School of Information and Electronics, Beijing Institute of Technology, Beijing. Her research interests include moving-target detection, parameter estimation and imaging.

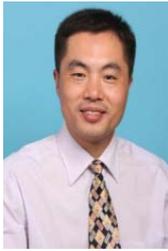
Wei Cui was born in Inner Mongolia Municipality, China in 1976. He received the Ph.D. degrees in electronics engineering from Beijing Institute of Technology in 2003. From 2003 to 2005, he worked as post-doctor in Radar Research Institute in Beijing Institute of Technology, where his research mainly concentrated on radar system and VLSI implementation of radar signal processing. Now, he worked as an professor and supervisor for doctorate students in Beijing Institute of Technology. His research interests include space target detection and localization, array signal processing, and VLSI design.

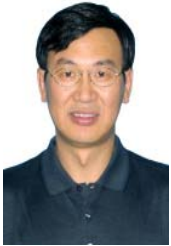
Siliang Wu was born in Anhui Province, China, in 1964. He received his Ph.D. degree from Harbin Institute of Technology in 1995 and then worked as a post-doctor in Radar Research Institute in Beijing Institute of Technology from 1996 to 1998. He is now a professor and supervisor for doctorate students in Beijing Institute of Technology and is a senior member of Chinese Institute of Electronics. His research interests include radar system and theory, satellite navigation and application of modern signal processing.